\documentclass[12pt]{article}
\usepackage{amsmath}
\usepackage{amsfonts}
\usepackage[utf8]{inputenc}
\usepackage{mathtools} 
\usepackage{amssymb}
\usepackage{geometry}
\usepackage{graphicx}
\usepackage{fancyhdr}
\usepackage{titlesec}
\usepackage{tikz-cd}
\usepackage{url}
\usepackage[T1]{fontenc} 
\usepackage{algorithm}
\usepackage{array}
\usepackage[caption=false,font=normalsize,labelfont=sf,textfont=sf]{subfig}
\usepackage{textcomp}
\usepackage{verbatim}
\usepackage{caption}
\usepackage{tabularray}
\usepackage{makecell}
\usepackage{algpseudocode}
\usepackage{tabularx}
\usepackage{rotating}
\usepackage{lipsum}
\usepackage{multirow}
\usepackage{lmodern}

\newtheorem{theorem}{Theorem}
\newtheorem{definition}{Definition}
\newtheorem{remark}{Remark}

\title{ALMA: a mathematics-driven approach for determining tuning parameters in generalized LASSO problems, with applications to MRI}
\author{$\textit{\textbf{Gianluca Giacchi}}^{4,6}$, $\textit{\textbf{Isidoros Iakovidis}}^{1, 6}$, $\textit{\textbf{Micah M. Murray}}^{2, 5}$, \\
$\textit{\textbf{Bastien Milani}}^{3, 5, 7}$, $\textit{\textbf{Benedetta Franceschiello}}^{3, 5, 7}$ \\
\small${}^{1}\textnormal{University of Bologna, Department of Mathematics, Bologna, Italy}$ \\ 
\small${}^{2}\textnormal{Lausanne University Hospital and University of Lausanne,}$ \\
\small$\textnormal{Department of Diagnostic and Interventional Radiology, Lausanne, Switzerland}^{2}$ \\ 
\small${}^{3}\textnormal{HES-SO Valais-Wallis School of Engineering,}$ \\ 
\small$\textnormal{Institute of Systems Engineering, Sion, Switzerland}^{3}$ \\
\small${}^{4}\textnormal{Università della Svizzera Italiana, Lugano, Switzerland}^{4}$ \\
\small${}^{5}\textnormal{The Sense Innovation and Research Center, Lausanne and Sion, Switzerland}^{5}$ \\
\small${}^{6}\textbf{These authors provided equal first-authorship contribution}^{6}$ \\
\small${}^{7}\textbf{These authors provided equal last-authorship contribution}^{7}$ \\
\small$\textnormal{Correspondence: Bastien Milani, Email: bastien.milani@hevs.ch}$}

\date{}

\begin{document}

\maketitle

\begin{abstract}

\noindent
\textbf{Summary:} ALMA (Algorithm for Lagrange Multipliers Approximation) is an iterative method for selecting tuning parameters in generalized LASSO problems for MRI reconstruction. Standardized approaches are often manual or heuristic; ALMA addresses this by approximating adaptatively Lagrange multipliers during reconstruction. On simulated MRI data, ALMA achieved mSSIM $\geq$ 0.99 and pSNR $\geq$ 40 dB, showing robust and near-optimal performance.

\noindent
\textbf{Purpose:} To introduce ALMA, a method for adaptatively computing tuning parameters in generalized LASSO problems during MRI reconstruction.

\noindent
\textbf{Methods:} We simulated MRI data using the Shepp-Logan phantom, testing ALMA across 450 reconstructions with varying undersampling (10–20\%) and noise levels (3–7\%). Image quality was assessed using mSSIM, pSNR, and CJV.

\noindent
\textbf{Results:} ALMA achieved average mSSIM of $0.9951 \pm 0.0041$, pSNR of $42.24 \pm 3.42$ dB, and CJV of $0.0367 \pm 0.0125$, converging in about $7.2 \pm 3$ iterations.

\noindent
\textbf{Conclusion:}ALMA offers a reliable, automated alternative to manual tuning, producing high-quality reconstructions and showing promise for in vivo MRI applications.
\end{abstract}


\section{Introduction}
Magnetic Resonance Imaging (MRI) is a non-invasive radiation-free technique that has become widespread over the last forty years. MRI exploits the magnetization of protons in water molecules of tissues to provide 3D anatomical images. To reduce the acquisition time, the MR signal is sampled below its Nyquist frequency, and the missing information is retrieved by solving numerically the generalized (unconstrained) LASSO (g-LASSO), a convex optimization problem in the form:
\begin{equation}
	\text{minimize} \quad \frac{1}{2}\Vert Ax-b\Vert_2^2+\frac{\lambda}{2}\Vert\Phi x\Vert_1, \qquad x\in \mathbb{C}^n,
\end{equation}
where $n$ is the dimensionality of the problem, $A\in\mathbb{C}^{m\times n}$ is a \textit{measurement operator}, $b\in\mathbb{C}^m$ is the \textit{measurement}, $\Phi\in\mathbb{C}^{N\times n}$ is a \textit{sparsity promoting transform}, and $\Vert\cdot\Vert_p$ denotes the $p$-(quasi-)norm on $\mathbb{C}^d$ ($0< p<\infty$):
\[
	\Vert y\Vert_p^p=\sum_{j=1}^d|\Re(y_j)|^p+|\Im(y_j)|^p, \qquad y\in\mathbb{R}^d.
\]
If $\Phi$ is the $n\times n$ identity matrix, g-LASSO is called LASSO. In the case of MRI, where the sampled signal is the Fourier transform of the proton density weighted by the effects of T1 and T2 time-relaxations, cf. \cite{mcrobbie2017mri}, $n$ is the number of voxels, $x$ is an image, $A=UFC$ encodes the undersampling operator $U$, the discrete Fourier transform (DFT) $F$ and $C$ is a complex matrix which encompasses the channel extension and the sensitivity mapping, $b$ is the noisy undersampled measurement and $\Phi$ can be the DFT, the discrete wavelet transform (DWT), the discrete cosine transform (DCT), or other sparsity promoting transforms, according to the situation, cf. \cite{lustig2008compressed}. 
The addendum $\Vert Ax-b\Vert_2^2$ measures the fidelity of the reconstruction, quantifying the distance between the measured data $b$ and the model $Ax$, whereas $\Vert\Phi x\Vert_1$ measures the sparsity of $\Phi x$. The parameter $\lambda>0$ acts as a trade-off, balancing the two contributions, and it is called tuning parameter. For small values of $\lambda$, the addendum $\lambda\Vert\Phi x\Vert_1$ becomes negligible and the image reconstructed via g-LASSO will be noisy and biased by the artifacts resulting from the missing information. On the contrary, large values of $\lambda$ force $\Vert\Phi x\Vert_1$ to be small, resulting in overly smoothed images with poor resolution (see fig.\ref{fig:figure2}B). 

The choice of this tuning parameter highly affects the quality of the reconstructions. However, there remains no generally effective procedure to provide well-performing tuning parameters. Consequently, the tuning parameter is chosen manually or heuristically, without following  standard protocols. An automatic procedure to detect optimal tuning parameters is therefore needed to heavily reduce the post-processing of MR acquired signals, to increase the reproducibility of MR reconstructions, and the adequacy of reconstructions, according to the specific clinical or research question at hand. Of note, the tuning parameters can also be chosen by some statistical model, but those need to be trained on good quality fully sampled datasets. While such training dataset exist for images of low dimension (3D or less), high-dimensional images (4D or more) cannot be completely sampled in practice. Finding a good tuning parameter to reconstruct such images therefore still rely on methods that are free of statistical models, such as ALMA. 

The mathematical reason why considering g-LASSO problems relates to the constrained LASSO problem (or quadratically constrained basis pursuit):
\begin{align}
	& \text{minimize} \qquad \Vert x\Vert_1 \\  
    & \text{subject to} \ x\in\mathbb{R}^n, \quad \Vert Ax-b\Vert_2\leq\eta, \nonumber
\end{align}
whose solution, if unique, is known to be $m$-sparse, where $m$ is the rank of the measurement matrix $A\in\mathbb{R}^{m\times n}$, cf. \cite{FR}. Under these assumptions, there exists a parameter $\lambda'\geq0$, related to the \emph{Lagrange multipliers}, such that the unconstrained LASSO has an $m$-sparse solution. For simplicity, we refer to $\lambda'$ as Lagrange multiplier. In broad terms, solving LASSO with tuning parameter $\lambda'$ provides sparse solutions. The question thus arises whether Lagrange multipliers play an analogous role for generalized LASSO problems, where $\Phi$ is a general sparsity promoting transform. To answer this question, we build an iterative algorithm, that we call ALMA (\emph{Algorithm for Lagrange Multipliers Approximation}), to provide an \emph{approximate Lagrange multipliers} (ALM) and reconstruct a MR image from undersampled data, artificially corrupted by Gaussian noise, according to the MRI sampling procedure, by solving g-LASSO with such obtained parameter. The quality of the reconstructions, along with the optimality of the tuning parameters, is measured by means of three metrics, quantifying the main aspects of an effective MRI reconstruction: the impact of noise and the impact of artifacts, see Section \ref{subsec:imagequalitymetrics} below. The purpose of this study is three-fold:
\begin{itemize}
	\item To define ALMA, an iterative procedure to compute ALMs, that can be easily generalized to more common versions of g-LASSO. 
	\item To understand the efficiency of ALMs as tuning parameters for TV weighted g-LASSO in the context of MRI.
	\item To assess the quality of reconstructions by means of image quality metrics.
\end{itemize}
Our results provide new insights into the impact of theoretical mathematical analysis on MR image reconstructions, shifting the focus from guessing tuning parameters to obtaining concrete estimates of noise energy.

\section{Methods}

\subsection{Theoretical model}

\subsubsection{Mathematical rationale}\label{sec:mrat}
The LASSO problem is fundamental in the recovery of sparse vectors. It stemmed by the contributions of F. Santosa and W.W. Symes \cite{santosa1986linear}, S. Chen and D. Donoho \cite{chen1994basis,donoho2006compressed}, and R. Tibshirani \cite{tibshirani1996regression}. Notably, within the domain of MRI, M. Lustig and colleagues \cite{lustig2008compressed,lustig2007sparse,ccukur2011signal,vasanawala2011practical,vasanawala2010improved} pioneered the applications of LASSO and g-LASSO. The purpose of this section is to motivate our conjecture that Lagrange multipliers could be used as effective tuning parameters for imaging reconstruction with g-LASSO. For the sake of completeness, we shall clarify the notion of sparsity comprehensively.\\

\begin{definition} A vector $x\in\mathbb{R}^n$ is $m$-sparse if $\Vert x\Vert_0=\#\{j:x_j\neq0\}\leq m$. When $m$ is irrelevant for the understanding, we say that $x$ is sparse.
\end{definition}

In the definition above, $\# A$ denotes the cardinality of the set $A$. A slightly modification of the proof \cite[Theorem 3.1]{FR} yields to the following result:

\begin{theorem}\label{thm1}
Let $A\in\mathbb{R}^{m\times n}$ be a measurement matrix and $\eta\geq0$. If the minimizer $x^\#$ of the constrained LASSO:
	\begin{align}\label{cLASSO}
		& \text{minimize} \qquad \Vert x\Vert_1 \\ 
        & \text{subject to} \ x\in\mathbb{R}^n, \ \Vert Ax-b\Vert_2\leq\eta \nonumber
	\end{align}
	is unique, then $x^\#$ is $\mbox{rk}(A)$-sparse, where $\mbox{rk}(A)$ denotes the rank of $A$.
\end{theorem}

\begin{remark}
	$A\in\mathbb{R}^{m\times n}$ is fundamental in Theorem \ref{thm1}, which fails if $A\in\mathbb{C}^{m\times n}$. For this reason, when working on $\mathbb{C}^d$, we consider its structure as the real vector space $\mathbb{R}^{2d}$ (see \cite{littleWolf}).
\end{remark}

LASSO has many equivalent formulations, where the equivalence notion is specified in \cite[Proposition 3.2]{FR}. We limit ourselves to delineate the relationship between the constrained LASSO \eqref{cLASSO} and its unconstrained counterpart:
\begin{equation}\label{ucLASSO}
	\text{minimize}\qquad \frac{1}{2}\Vert x\Vert_1+\frac{\lambda}{2}\Vert Ax-b\Vert_2^2.
\end{equation}

\begin{theorem}\label{thm2} Let $A\in\mathbb{R}^{m\times n}$, $b\in\mathbb{R}^m$ and $\eta>0$. Let $x^\#$ be a minimizer of the constrained LASSO \eqref{cLASSO}. Then, there exists $\lambda'\geq0$ such that $x^\#$ is also a minimizer of \eqref{ucLASSO} with $\lambda=\lambda'$.
Conversely, if $x^\#$ is a minimizer of \eqref{ucLASSO}, there exists $\eta'\geq0$ such that $x^\#$ is also a minimizer of \eqref{cLASSO} with $\eta=\eta'$.
\end{theorem}

Consequently, solving \eqref{ucLASSO} with the corresponding Lagrange multiplier provides a $\mbox{rk}(A)$-sparse solution. Theorem \ref{thm2} holds also for g-LASSO:
\begin{equation}\label{g-LASSO}
	\text{minimize} \qquad \frac{1}{2}\Vert Ax-b\Vert_2^2+\frac{\lambda}{2}\Vert \Phi x\Vert_1
\end{equation}
Note that the rescaling by the factor $1/2$ is due to a choice of implementation of the reconstruction, but it is irrelevant to the analysis of \eqref{g-LASSO} from a mathematical point of view. Also, observe that \eqref{g-LASSO} with $\lambda\neq0$ is equivalent to:
\begin{equation}\label{g-LASSO2}
	\text{minimize} \qquad \frac{1}{2}\Vert \Phi x\Vert_1+\frac{\lambda}{2}\Vert Ax-b\Vert_2^2,
\end{equation}
where the equivalence follows by choosing $\lambda^\ast=\lambda^{-1}$. 

\subsubsection{Construction of Lagrange multipliers}
\begin{figure}[ht]	\begin{center}\includegraphics[width=0.45\textwidth]{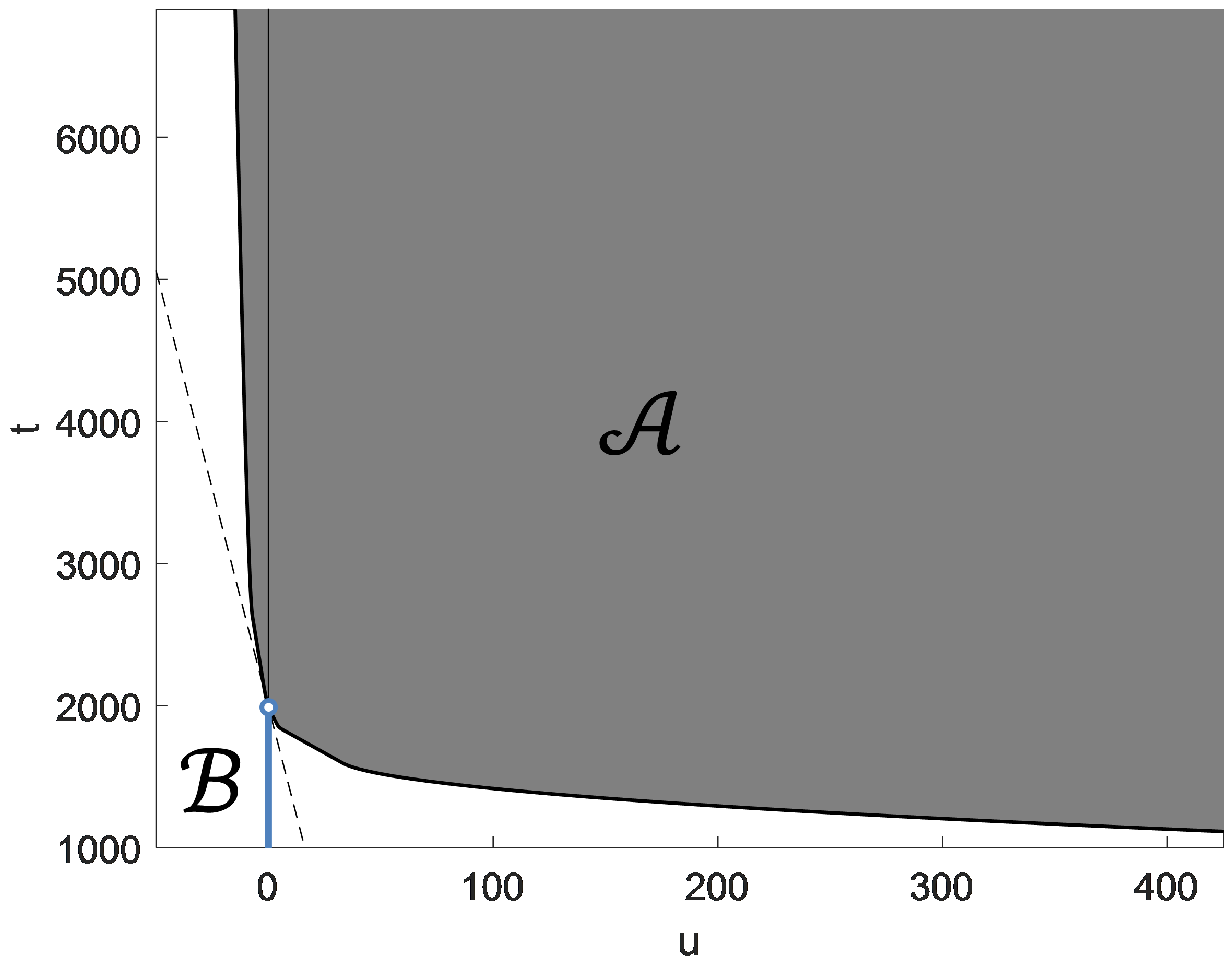}\end{center}
	\caption{A graphic representation of the sets $\mathcal{A}$ and $\mathcal{B}$, and the separating hyperplane (dashed line). Observe that $\mathcal{A}$ is an epigraph and $\mathcal{B}$ is an open lower half-line. The closure of $\mathcal{B}$ intersects the boundary of $\mathcal{A}$.}
	\label{fig:setsAB}
\end{figure}

The construction of a Lagrange multiplier is detailed in the proof of \cite[Theorem 4.8]{boyd2004convex}, which uses Hahn-Banach theorem to find a hyperplane that separates two convex sets (see fig.\ref{fig:setsAB}), which for \eqref{g-LASSO2} read as the epigraph 
\begin{align}
\mathcal{A}=\Big\{ & (u,t)\in\mathbb{R}^2: \\
                   & u\geq\frac{1}{2}\Vert Ax-b\Vert_2^2-\frac{\eta^2}{2}, \ t\geq\frac{1}{2}\Vert\Phi x\Vert_1, \nonumber \\  
                   & \text{for some } x\in\mathbb{R}^n\Big\} \nonumber
\end{align}
and the lower half-line 
\begin{equation}
\mathcal{B}=\Big\{(0,t)\in\mathbb{R}^2:t<p^\ast\Big\}
\end{equation} 
where 
\begin{equation}
p^\ast=\min\{\Vert\Phi x\Vert_1 : \Vert Ax-b\Vert_2\leq\eta\}
\end{equation}
. 

In the particular case of g-LASSO, separating hyperplanes are lines, and if $t=mu+q$ is any separating line, then a Lagrange multiplier can be chosen as $\lambda^\ast=-m$ or, equivalently, $\lambda=-1/m$. Let us observe that set $\mathcal{B}$ serves no essential purpose in determining a separating line, as it merely constitutes a lower half-line intersecting $\mathcal{A}$ at its boundary. Additionally, when the lower boundary of $\mathcal{A}$ is $\mathcal{C}^1$ regular in a neighborhood of $u=0$, a numerical approach enables the identification of a separating line by delineating $\mathcal{A}$ and computing the tangent at $0$ along the graph of its boundary.

\subsubsection{Main challenges}
A direct application of the theory described so far for deriving tuning parameters poses three main challenges.
	\begin{itemize}
	\item \textbf{Necessity of numerical methods:} in the vast majority of convex optimization problems, the expression of the dependence $\lambda=\lambda(x^\# )$ or even $\lambda=\lambda(\eta)$, is a challenging task. We refer to \cite{giacchi2024determination} for examples of weighted LASSO problems where this relation is instead explicit. Hence, finding a Lagrange multiplier with the construction in \cite{boyd2004convex} requires a separating line to be found numerically.
	\item \textbf{Unknown constraints:} theoretically, the numerical machinery described above necessitates prior knowledge of $\eta=\Vert Ax^\#-b\Vert_2$, where $x^\#$ is the solution of g-LASSO.
	Consequently, the cylinder $\Gamma_\eta=\{x:\Vert Ax-b\Vert_2\leq\eta\}$ is not only unknown, but it depends heavily on the outcome of g-LASSO. 
	\item \textbf{Dimensionality:} outlining $\mathcal{A}$ or its boundary necessitates plotting $\infty^n$ points in $\mathbb{R}^2$, rendering it an impractical endeavor. 
	\end{itemize}

While acknowledging the necessity of numerical analysis, we may resort to an escamotage to overcome the two remaining challenges. However, this entails abandoning the pursuit of exact Lagrange multipliers in favor of approximations. The constraint bound $\eta$ depends intrinsically on the solution $x^\#$ of g-LASSO, which in general differs from the ground truth $f$ which, in the case of our experimental setting, consists of the Shepp-Logan phantom (see fig.\ref{fig:radial}C ).

% Following our discution, I would rephrase it. 
% As a solution, we substitute $\Vert Ax^\#-b\Vert_2$ with $\eta=\Vert % Af-b\Vert_2$, where, $f$ is the Shepp-Logan brain phantom.
% For example (you can also rephrase) : 

Defining a constrained bound $\eta$ induces a solution $x^\#$ and reversely, a given solution $x^\#$ sets the constraint bound $\eta$. In practice the solution $x^\#$ is of course unknown and it is a subtle task to choose $\eta$ appropriately so that the solution $x^\#$ is, hopefully, as close as possible to the ground truth. Choosing $\eta$ is equivalent to choosing the allowed amount of error on the raw data. It follows that $\eta$ must be at least as large as the noise amplitude. Other sources of error can also add to the noise so that in general $\eta$ may be larger than the noise amplitude. This being said, in the present study, we chose to set $\eta$ to be equal to the norm of the added noise 

\[\eta=\Vert  Af-b\Vert_2 = \Vert  \varepsilon \Vert_2.\]

This situation is still far from concrete, since $\Vert  \varepsilon \Vert_2$ is not known in the practice. However, we note that the purpose of the present study is to present an ideal situation to test if it is possible to use approximations of Lagrange multipliers as well-performing tuning parameters for generalized LASSO problems, shifting the focus from selecting the tuning parameter to estimating the noise energy. 

Consequently, the tuning parameter returned by ALMA serves as an \emph{approximate Lagrange multiplier}, rather than the exact one. Achieving this approximation involves outlining the corresponding epigraph $\mathcal{A}$. The essence of ALMA lies in a technicality allowing for the approximation of $\mathcal{A}$ by sketching infinitely many of its points simultaneously, as outlined below.

\subsubsection{Comparison with the L-curve parameter}
In this work, we compare the performance of ALMA with the L-curve method, a classical approach to determining and computing a suitable value for the regularization parameter $\lambda$ in \eqref{g-LASSO}. It was initially introduced by Lawson and Hanson in \cite{Lawson1976SolvingLS} and it consists of considering the curve:
\begin{equation}\label{L-curve}
	L=\{(\log\Vert Ax_\lambda-b\Vert_2^2,\log\Vert\Phi x_\lambda\Vert_1), \quad \lambda>0\}, 
\end{equation}
where $x_\lambda$ is the solution of \eqref{g-LASSO} with respect to the corresponding parameter $\lambda$. Hansen and O’Leary in \cite{Han1, Han2} proposed that the parameter corresponding to the corner of the graph \eqref{L-curve} provides the best trade-off between $\Vert Ax - b\Vert_2^2$ and $ \Vert \Phi x\Vert_1$.

\subsubsection{The MRI model}
%\begin{figure*}
%	\begin{center}\includegraphics[width=\textwidth]{Sampling.png}\end{center}
%	\caption{Full-Cartesian 3D sampling.}
%	\label{fig:sampling}
%\end{figure*}
For the sake of concreteness, we simulate the reconstruction of a MR signal. Let us spend a few words about how g-LASSO is used, and why the correct choice of $\lambda$ is fundamental, in the context of MRI. Roughly speaking, the (inverse) spatial Fourier transform of the MR signal is the anatomical image of a tissue, which is supported in a box $[-L_x/2,L_x/2]\times[-L_y/2,L_y/2]\times[-L_z/2,L_z/2]$. By Shannon’s theorem, full Cartesian (uniform) sampling consists of sampling $n_x$ points in the $k_x$ direction, $n_y$ points along the $k_y$ direction and $n_z$ points along the $k_z$ direction of the k-space, where:
\[
	\Delta x =\frac{L_x}{n_x}, \ \Delta y = \frac{L_y}{n_y}, \ \Delta z = \frac{L_y}{n_z},
\]

and

\[
	\Delta k_x =\frac{1}{L_x}, \ \Delta k_y = \frac{1}{L_y}, \ \Delta k_z = \frac{1}{L_z},
\]
We mention McRobbie et.al \cite{mcrobbie2017mri} as reference therein. 

For various reasons, sampling the MRI signal at its Nyquist frequency poses several challenges. These challenges necessitate techniques that can accurately reconstruct the MRI signal from samples taken below the Nyquist frequency. First, in our 2D experiments the dimensionality of the problem is of the order of $n=n_xn_y=384^2\sim10^5$. For a 3D image, it increases to $n=n_xn_yn_z\sim10^8$, making the processing of the MRI signal extremely time-consuming and computationally expensive. Secondly, MRI requires patients to remain still throughout the entire acquisition process, making it challenging to image moving organs and to perform scans on patients with conditions such as movement disorders. Furthermore, MRI is extremely expensive, and reducing the amount of information to be acquired can lead to significant cost savings. 

In light of these reasons, CS offers a valuable solution for reducing acquisition time. However, there are situations where CS transcends being merely advantageous: it becomes imperative. Take, for example, 3D-CINE MRI, where sampling a full-Cartesian grid would be excessively time-consuming to the extent that achieving full sampling becomes practically infeasible. The application of CS to MRI is called CS-MRI. Since the MR signal is known to be sparse with respect to the DFT, DWT, and other sparsity promoting transforms, cf. \cite{lustig2008compressed}, g-LASSO is used for the purpose. In this work, we use the Shepp-Logan phantom and corrupt the data with artificial noise. The Shepp-Logan phantom is piecewise constant and, therefore, its gradient is sparse. For this reason, we use the discrete gradient as sparsity promoting transform:
\begin{equation}\label{TV-2}
	D = \begin{pmatrix}-1 & 1 & 0 & \ldots & 0 & 0\\
	0 & -1 & 1 & \ldots & 0 & 0\\
	\vdots & \vdots & \vdots & \ddots & \vdots & \vdots\\
	0 & 0 & 0 & \ldots & -1 & 1
	\end{pmatrix},
\end{equation}
and we set $TV(x)=\Vert D x\Vert_1$ (discrete anisotropic spatial total variation).

\subsection{Experimental design}
The experiments were conducted using MATLAB\textunderscore R2023b. The Monalisa toolbox was used for MRI reconstructions.
\subsubsection{The simulated MR signal}
To simulate the acquisition of an MR signal, we considered the Shepp-Logan brain phantom $f\in\mathbb{R}^{384\times384}$, simulate coil sensitivity, undersampling, and Gaussian noise. MR data is sampled by a certain number $nCh$ of coils simultaneously (parallel imaging). Let us delve into a detailed exposition of the data simulation, briefly summarized in the previous lines. 
	\begin{enumerate}
	\item \textbf{Channel extension and acquisition across coils:} $f$ is replicated $nCh=8$ times and each replica $f_j\in\mathbb{R}^{384\times384}$, $j=1,\ldots,nCh$, is pointwise multiplied by a simulated coil sensitivity matrix $C_j\in\mathbb{C}^{384\times 384}$:
	$$f_k (i,j)=C_k (i,j)f(i,j),$$ $k=1,\ldots,nCh$, $i,j=1,\ldots,384$ (fig. \ref{fig:figure2}A). 
    
\begin{figure*}
	\begin{center}
		\includegraphics[width=\textwidth]{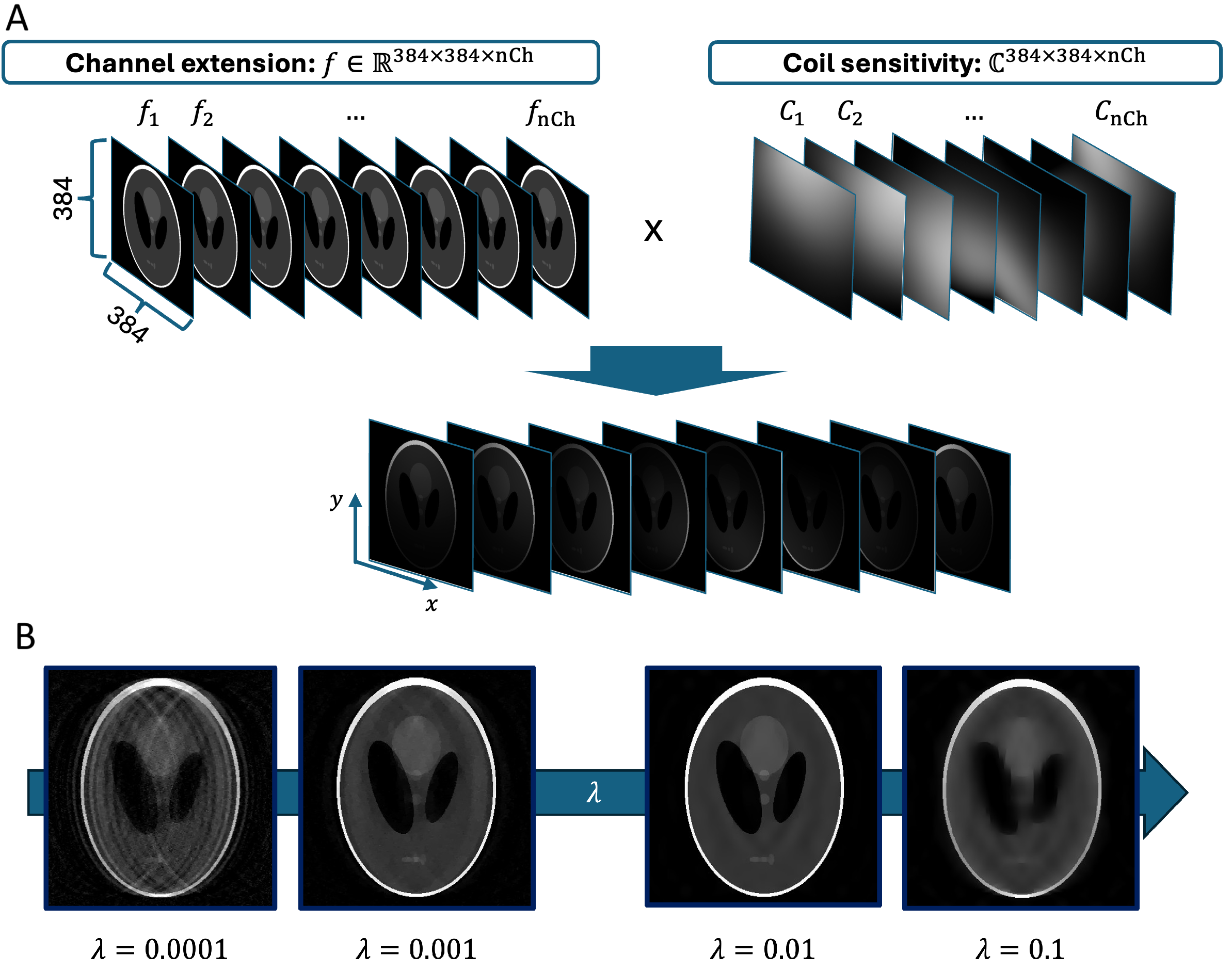}
		\caption{A: Simulated channel extension and pointwise multiplication by the coil-sensitivities. B: Reconstructions with different tuning parameters. For small values of $\lambda$, the reconstructed image exhibits pronounced artifacts. Large values of $\lambda$ produce oversmoothing. }
		\label{fig:figure2}
	\end{center}
\end{figure*}

	\item \textbf{Fourier transform:} the spatial discrete Fourier transform of each $f_j$ is computed. 
	\item \textbf{Undersampling:} full Cartesian sampling consists of sampling $nLines=384$ lines evenly spaced, whereas undersampling entails sampling only a certain fraction of these 384 lines, which we denote by ${UR}_\%$. In our study, we tested undersampling rates of $10\%$, $15\%$ and $20\%$, that is ${UR}_\%\in\{10/100,15/100,20/100\}$. For a fixed ${UR}_\%$, the sampling trajectory comprises $n({UR}_\%)=\lceil nLines\cdot {UP}_\% \rceil$ lines. The $30\%$ of the $n({UR}_\% )$ lines are used to sample the center of the k-space at the Nyquist frequency. Specifically, $\lceil n({UR}_\%)\cdot 30/100\rceil$ lines sample the center of the k-space, while the remaining lines sample the periphery of the k-space following a normal distribution $\mathcal{N}(\mu,\sigma^2 )$ with $\mu=nLines/2 +1$ and $\sigma^2=nLines\cdot {UR}_\%$. The Fourier transform of each coil-image is sampled according to the Cartesian undersampled trajectory established before. Consequently, the resulting simulated MR data consists of a tensor $y\in Y := \mathbb{C}^{(nLines^2\cdot n({UR}_\% ))\times nCh}$. 
	\item \textbf{Noise corruption:} we tested ALMA under three distinct noise levels. Specifically, the simulated MR data is given by: $b=y+\varepsilon$, where $\varepsilon\in Y$ and $\Re(\varepsilon_{i,j}),\Im(\varepsilon_{i,j} )\sim \mathcal{N}(0,\sigma^2 )$ ($i=1,\ldots,nLines\cdot n({UR}_\% $) and $j=1,\ldots,nCh$). We refer to $\sigma^2$ as to noise level, which is computed as $\sigma^2=\Vert y\Vert\cdot {NL}_\%$, where in our experiments ${NL}_\%\in\{3/100,5/100,7/100\}$. For instance, the terminology $3\%$ noise means that we are considering ${NL}_\%=3/100$.
\end{enumerate}

\subsubsection{ALMA}
\begin{figure*}
\begin{center}\includegraphics[width=\textwidth]{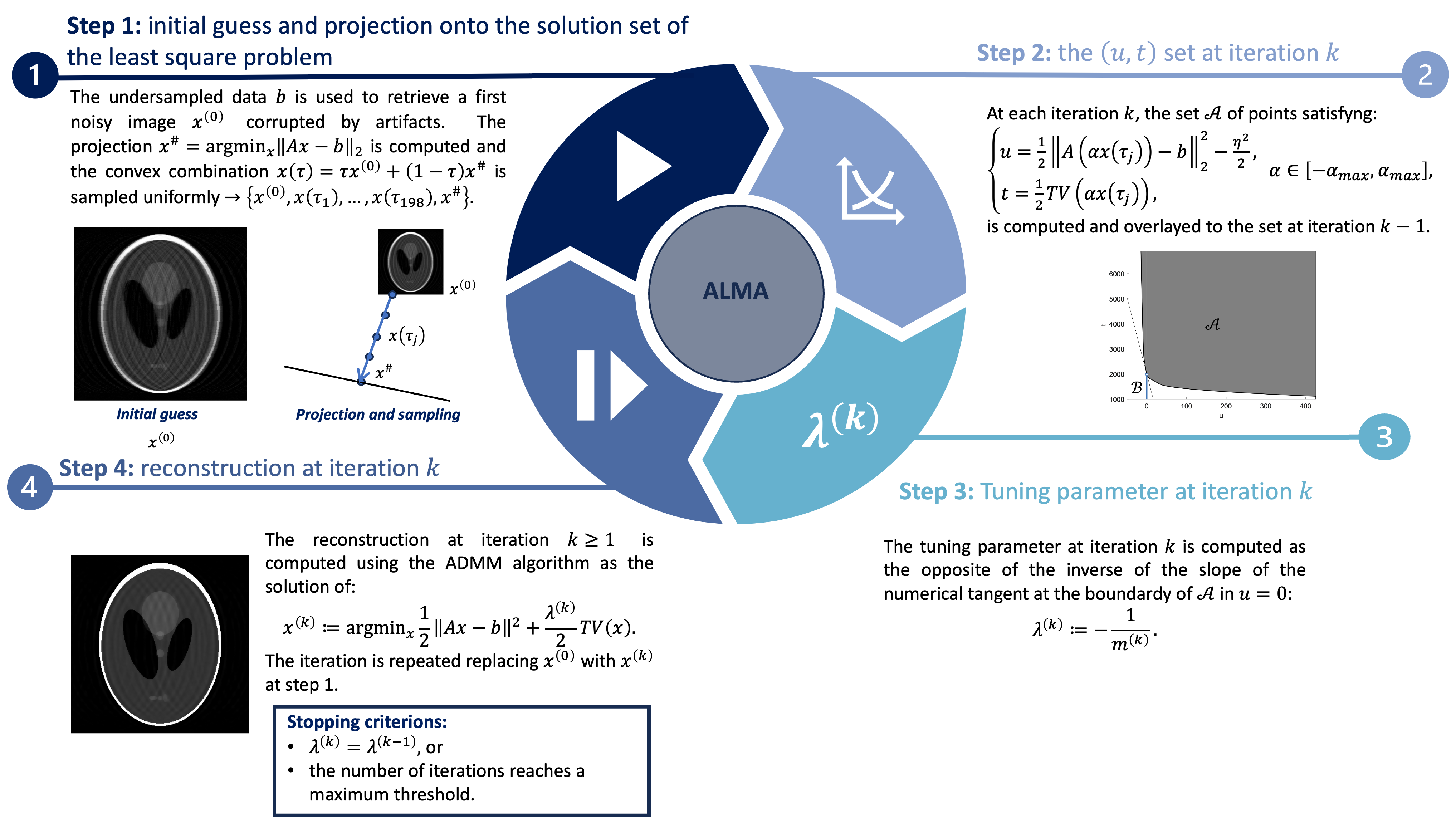}\end{center}
	\caption{Schematic representation of ALMA.}
	\label{fig:ALMA}
\end{figure*}
ALMA is synthesised in fig. \ref{fig:ALMA} and schematized in Algorithm \ref{alg:cap}, here we limit to comment how ALMA computes a ALM.
\begin{algorithm}
\caption{ALMA}\label{alg:cap}
\begin{algorithmic}
\Require: maximal number of iterations: $n_{max}$.
\Require: $A\in\mathbb{C}^{m\times n}$ and $b\in\mathbb{C}^m$.
\Require: $x^\#$ ground-truth and set $\eta=\Vert{Ax^\#-b}\Vert_2$.
\Require: $x^{(0)}=\arg\min_{x\in\mathbb{C}^n}\Vert{Ax-b}\Vert_2$.
\While{$n \leq n_{max}$}\\
1. Project $x^{(n-1)}$ onto the solution set of the least-square problem, call $x^{(n-1)}_{proj}$ the projection.\\
2. Consider the convex combination $x_\tau^{(n)}=\tau x^{(n-1)}+(1-\tau) x^{(n-1)}_{proj}$ ($0\leq\tau\leq1$).\\
3. Consider a partition $\tau_1=0,\ldots,\tau_{200}=1$ of $[0,1]$ and the related $x_{\tau_j}^{(n)}$.\\
\While {$1\leq j\leq 200$}
Plot the set $\mathcal{A}$ of points $(u,t)\in\mathbb{C}^2$ in the form
\[
\begin{cases}
u = \frac{1}{2}\Vert{A(\alpha_k x_{\tau_j}^{(n)})-b}\Vert_2^2-\frac{\eta^2}{2},\\
t = \frac{1}{2}TV(\alpha_k x_{\tau_j}^{(n)}),
\end{cases}
\]
where $\alpha_k\in[-\alpha_{max},\alpha_{max}]$ is an equally spaced sequence ($k=1,\ldots,k_{max}$),
\[
\alpha_{max}=\frac{|b^T Ax_{\tau_j}^{(n)} |}{\Vert{Ax_{\tau_j}^{(n)}}\Vert_2^2}.
\]
\EndWhile \\
4. Compute the slope $m^{(n)}$ of the tangent to the lower boundary of $\mathcal{A}$, in $u=0$. Set $\lambda^{(n)}=-1/m^{(n)}$.\\
5. Solve 
\[
    \arg\min_{x\in\mathbb{C}^n}\frac{1}{2}\Vert Ax-b\Vert_2^2+\frac{\lambda^{(n)}}{2}TV(x)
\]
with the ADMM algorithm. Call $x^{(n)}$ the solution.\\
\noindent
\If{ $n>1$ and $\lambda^{(n)}=\lambda^{(n-1)}$ }
Break
\EndIf
\EndWhile \\
\textbf{Result:} $x_{out}=x^{(r)}$, where $r$ is the number of iterations at the end of \textbf{while}.
\end{algorithmic}
\end{algorithm}
Note that the scalar product $a^T b$, $a,b\in\mathbb{R}^n$, must be replaced with $Re(a^T b)$ when $a,b\in\mathbb{C}^n$, where $Re$ denotes the \textit{real part} \cite{littleWolf}. Moreover, for non-Cartesian MRI trajectories, the scalar product that shall be considered is $\langle a, b \rangle=Re(a^T H b)$, with $H$ Hermitian positive-definite matrix encoding the non-constant density of the k-space sampling. 

In the previous paragraphs we observed that finding an ALM is a matter of tracing the tangent line in 0 to the epigraph
\begin{align*}
\mathcal{A}=\{  & (u,t)\in\mathbb{R}^2 :  \\
                & u\geq\Vert Ax-b\Vert_2^2/2-\eta^2/2, \ t\geq TV(x)/2, \\
                & \ \text{for \ some} \ x\in\mathbb{C}^n\}
\end{align*}
where $\eta=\Vert Af-b\Vert_2$, being $f\in\mathbb{R}^{384\times 384}$ the $384\times 384$ Shepp-Logan phantom. We pointed out that outlining $\mathcal{A}$ is technically difficult due to the unfeasible dimensionality: in principle, for every $x\in\mathbb{C}^n$ ($n=384^2$), once the point 
\begin{equation*}
(u,t)=(u(x),t(x))=1/2\cdot( \Vert Ax-b\Vert_2^2-\eta^2,  TV(x)) 
\end{equation*}
is computed, one has that all the points of the first quadrant centered in $(u(x),t(x))$ belong to $\mathcal{A}$. Clearly, it would be enough to compute $(u(x),t(x))$ only for $x\in\mathbb{C}^n$ such that $(u(x),t(x))$ belongs to the boundary $\partial\mathcal{A}$ of $\mathcal{A}$, but we do not have access to those points. On top of that, computing $(u(x),t(x))$ for a fixed $x$ is computationally expensive, because of the measurement operator $A$. However, a meaningful family of points $(u(x),t(x))$ that would be enough to approximate the tangent line in 0 to its boundary can be found as follows:
\begin{enumerate}
	\item  Choose $x\in\mathbb{C}^n$ so that the corresponding $(u(x),t(x))$ is as far to the left of $\mathcal{A}$ as possible. Clearly, this task is accomplished by any minimizer of $\phi(x)=\frac{1}{2}\Vert Ax-b\Vert_2^2$. Let us call this point $x^\#$.
	\item  Choose another point, for instance, the reconstructed image obtained by the gridded reconstruction of the noisy undersampled data, $b$. Let us call this point $x^{(0)}$. 
	\item	 Consider the segment that joins $x^\#$ to $x^{(0)}$ in the image domain and sample it at a rate decided a priori, e.g. 201 uniformly spaced samples. 
	\item Let $x$ be one of this samples. Compute $\Vert Ax\Vert_2^2$, $b^T Ax$ and $TV(x)$.
	\item The curve $\gamma_x$ parametrized by $\alpha\in\mathbb{R}$ as:
	\begin{align*}
		\gamma_x&(\alpha)=(u(\alpha x),t(\alpha x))\\
		&=\frac{1}{2}\Big(\alpha\Vert Ax\Vert_2^2-2\alpha b^TAx+\Vert b\Vert_2^2-\eta^2,|\alpha|TV(x)\Big)
	\end{align*}
	consists of a couple of branches of parabolas. The parameters $\alpha_1$ and $\alpha_2$ such that $\gamma_x(\alpha_1 )$ and $\gamma_x(\alpha_2 )$ are the vertices of these parabolas can be computed explicitly by the expression of $\gamma_x(\alpha)$.
	\item Let $\alpha_{max}=\max\{|\alpha_1|,|\alpha_2|\}$. Compute and plot $\gamma_x(\alpha)$ for $\alpha\in[-\alpha_{max},\alpha_{max }]$. Since $\Vert Ax\Vert_2^2$, $b^T Ax $ and $TV(x)$ have been computed in 4., the computational cost of this operation is low.
	\item Repeat the procedure for every $x$ belonging to the segment that joins $x^\#$ to $x^{(0)}$.
	\item Since $\mathcal{A}$ is known to be convex, compute the convex hull of the outlined points.
\end{enumerate}
The convex hull at the end of 8. is the approximation of $\mathcal{A}$ at iteration 1. 
\begin{enumerate}
\setcounter{enumi}{8}
	\item Choose $\lambda^{(1)}=-1/m^{(1)}$, where $m^{(1)}$ is the slope of the tangent and reconstruct an image using $\lambda^{(1)}$ as tuning parameter for TV-LASSO. Call this image $x^{(1)}$. Note that this step consist in performing a reconstruction. This is why ALMA can be considered as an iterative reconstruciton while adapting $\lambda$ until convergence.   
\end{enumerate}
Indicatively, $x^{(1)}$ has the advantage of being more regular than $x^{(0)}$, that is $TV(x^{(1)} )\leq TV(x^{(0)} )$.
 \begin{enumerate}
\setcounter{enumi}{9}
	\item Repeat steps 1)-9) replacing $x^{(0)}$ with $x^{(1)}$, to outline points of $\mathcal{A}$ that are narrower with respect to the points outlined in steps 1)-7). Overlaying these new points to the ones already outlined in 7), improves the approximation of $\mathcal{A}$.
	\item  Compute the slope $m^{(2)}$ of the new tangent in 0 to the convex boundary of $\mathcal{A}$ and define $\lambda^{(2)}=-1/m^{(2)}$. 
\end{enumerate}

\subsubsection{Image quality metrics}\label{subsec:imagequalitymetrics}
We measured quantitatively the quality of the output of ALMA by means of three metrics: the mSSIM, the pSNR and the CJV. %We chose mSSIM as an overall quality measure of the reconstruction, the pSNR as a measure of noise and CJV as a measure of the biased due to the artifacts. 
\begin{itemize}
\item The mSSIM is an extension of the structural similarity index, designed to assess the quality of reconstructions across various scales in a manner that approximates human perception,\cite{wang2003multiscale, deligiannidis2014emerging}. It compares the brightness, contrast, and structural details of reconstructions with ground truth images, assigning values on a scale from 0 to 1, where a score of 1 indicates optimal similarity. Good quality reconstructions typically correspond to mSSIM values of $\geq 0.9$. In the present work, the mSSIM is computed via the command \texttt{multissim(I,Iref)}, where \texttt{Iref} is the reference image (the Shepp-Logan brain phantom) and \texttt{I} is the image to be assessed.
\item The pSNR quantifies noise corruption of compressed images, independently on the quality as perceived by human vision  and good visual quality requires pSNR to be at least 30dB, \cite{adcock2021compressive}.
\item CJV measures the presence of intensity non-uniformity (INU) artifacts in MRI \cite{likar2001retrospective,ganzetti2016intensity}. In the current paper, we use it as a measure of artifact bias in reconstructions. Lower values of CJV indicate better quality of MR images in terms of artifacts. CJV is defined based on the intensity difference between grey and white matter, as inspired by Jaroudi et.al. \cite{jaroudi2019numerical}.
\end{itemize}

The notation $mSSIM(\lambda)$ stands for the mSSIM of the reconstruction obtained by solving:
\begin{equation}\label{TV-LASSO}
	\arg\min_{x\in\mathbb{C}^n}\frac{1}{2}\Vert Ax-b\Vert_2^2+\frac{\lambda}{2}TV(x).
\end{equation}
Analogous notations for $pSNR(\lambda)$ and $CJV(\lambda)$. The mSSIM takes values in $[0,1]$ and optimality corresponds to $mSSIM(\lambda)=1$ (maximum). The pSNR and the CJV take values in $[0,+\infty)$ and are optimized in correspondence of their maxima and their minima respectively. Good quality with respect to mSSIM corresponds to $mSSIM \geq 0.9$, good quality with respect to pSNR corresponds to $pSNR \geq$ 30dB. Assessing good quality for CJV is harder, since, differently from the mSSIM and the pSNR, the CJV is optimized in correspondence of its minima $\arg\min(CJV(\lambda))$, whereas $\max(CJV(\lambda))$ potentially grows to infinity. For these reasons, we considered good quality with respect to CJV as:
\begin{equation}\begin{split}\label{CJVthre}
	CJV(\lambda)&\leq\min(CJV)+\frac{\max(CJV)-\min(CJV)}{10}\\
	&=0.0493\approx0.05,
\end{split}\end{equation}
where $\max(CJV)=\max\{CJV(\lambda):0\leq\lambda\leq2\}$ and $\min(CJV)=\min\{CJV(\lambda):\lambda\geq0\}$, analogous notations are used for the minima and the maxima of mSSIM and pSNR. 
\subsubsection{Data analysis}
	The mSSIM, pSNR and CJV serve as quantitative measures for assessing the quality of reconstructions. Additionally, the (unique, in our experiments) tuning parameters $\lambda_{mSSIM},\lambda_{pSNR}$ and $\lambda_{CJV}$ which optimize each metric can be considered. Specifically, $\lambda_{mSSIM}=\arg\max\{mSSIM(\lambda):\lambda\geq0\}$, $\lambda_{pSNR}=\arg\max\{pSNR(\lambda):\lambda\geq0\}$  and $\lambda_{CJV}=\arg\min\{CJV(\lambda):\lambda\geq0\}$. The ratios $\lambda_{mSSIM}/\lambda_{ALM}$, $\lambda_{pSNR}/\lambda_{ALM}$ and $\lambda_{CJV}/\lambda_{ALM}$ express the distance between optimal tuning parameters (with respect to the metrics) and ALMs in terms of the orders of magnitude of the corresponding ratios. We term the ratio $\lambda_{mSSIM}/\lambda$ the magnitude ratio corresponding to $\lambda$ (with respect to mSSIM), and similar notations are reserved for the other metrics. For every pair $({UR}_\%,{NL}_\% )$ and every reconstruction, we calculate the mSSIM, pSNR and CJV, alongside their corresponding magnitude rations. Within each fixed pair $({UR}_\%,{NL}_\% )$, we employ violin plots to illustrate the distributions of mSSIM, pSNR and CJV across the 50 runs, for analysis. Additionally, shaded error bars for the functions $mSSIM(\lambda/\lambda_{ALM})$, $pSNR(\lambda/\lambda_{ALM})$ and $CJV(\lambda/\lambda_{ALM})$  are incorporated to evaluate the optimality of ALMs as tuning parameters. A value of $\lambda_{mSSIM}/\lambda_{ALM}=1$ signifies that $\lambda_{ALM}$ is the optimal tuning parameter with respect to mSSIM. Similar interpretations apply for the other magnitude ratios. For a comparison, we compute the L-curve tuning parameter $\lambda_L$, and execute reconstructions using TV-LASSO with this parameter. Subsequently, we compute $mSSIM(\lambda_L )$, $ pSNR(\lambda_L )$ and $CJV(\lambda_L )$, along with the corresponding magnitude ratios pertaining to $\lambda_L$ with respect to these three metrics. 

\subsubsection{The dataset}
The dataset is obtained reconstructing an image for a fixed pair $({UR}_\%,{NL}_\% )$ 50 times (number of runs), to ensure statistical robustness with respect to the sampling randomisation. The dataset consists of $3\times 3\times 50$ reconstructions.

\section{Results}
	Violin plots for the three metrics across noise levels, undersampling rates and number of runs are displayed in figure \ref{fig:Tabellona}. The average and standard deviation values of these metrics, as well as their average magnitude ratios, are reported in figure \ref{fig:table1}. 
\begingroup
\setlength{\tabcolsep}{20pt} % Default value: 6pt
\renewcommand{\arraystretch}{10} % Default value: 1

		\begin{figure*}
	\begin{center}\includegraphics[width=\textwidth]{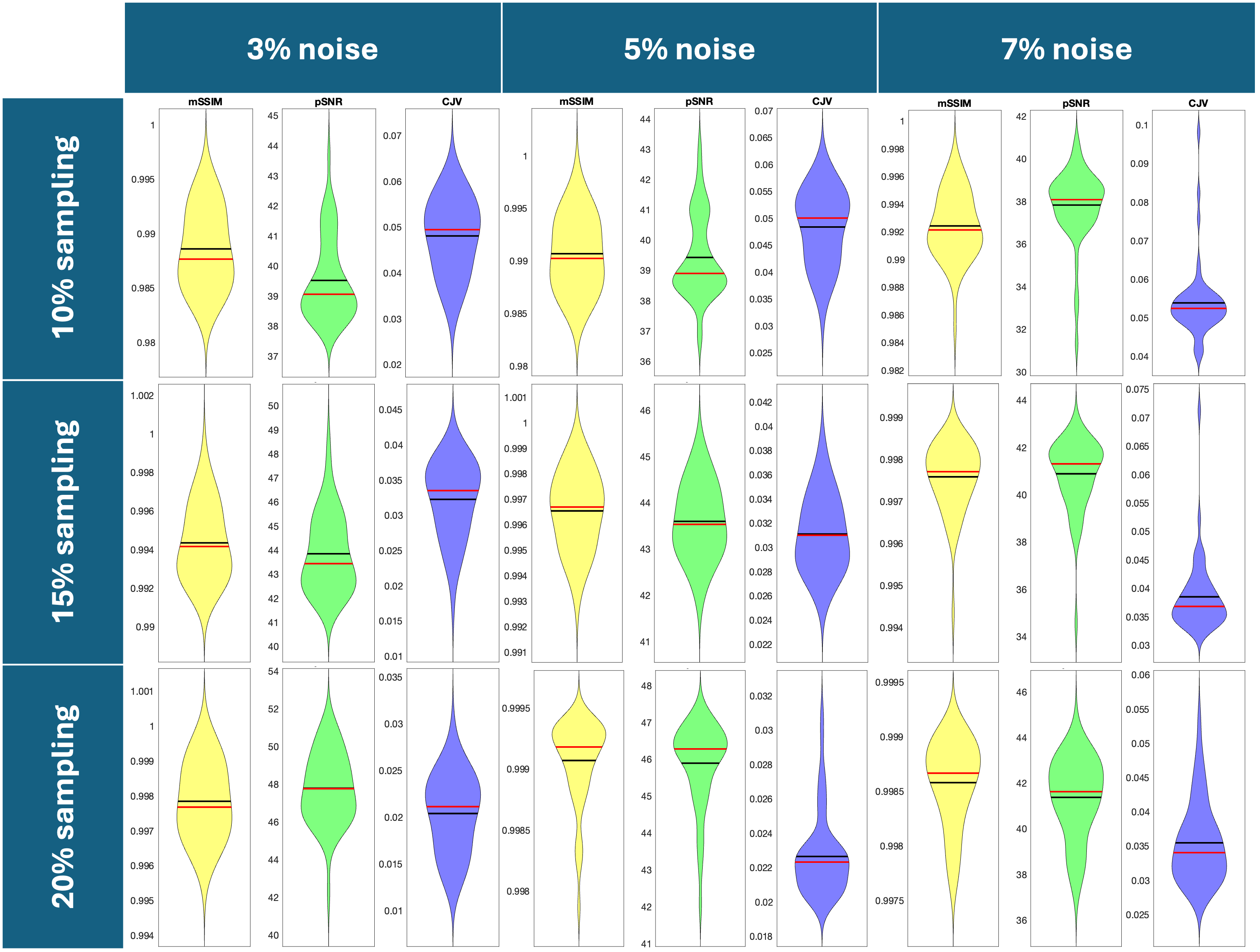}\end{center}
	\caption{Violin plots of the three metrics across noise levels and undersampling rates. Yellow violins correspond to mSSIM, green violins correspond to pSNR and purple violins correspond to CJV. The black lines correspond to the means and the red lines correspond to the medians.}
	\label{fig:Tabellona}
\end{figure*}

\subsection{Analysis of convergence}
    
\begin{figure*}
	\begin{center} 
    \includegraphics[width=0.95\textwidth]{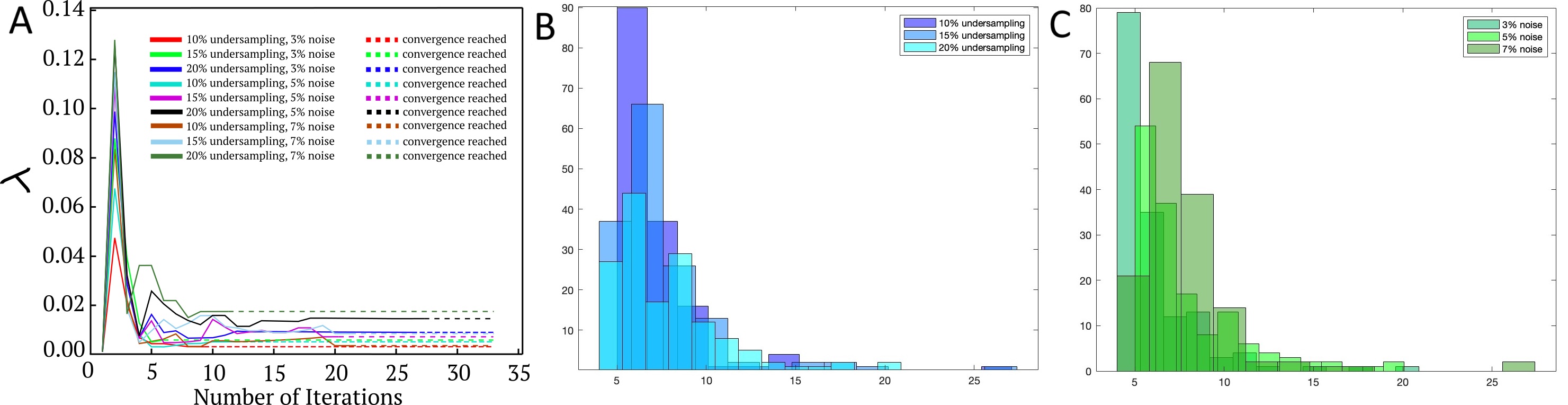}
	\caption{Convergence analysis. Subfigure A shows examples of the convergence of ALMA across noise levels and undersampling rates. Each graph corresponds to a fixed couple: noise level and undersampling rate, as illustrated in the legend, showing the iterative progression of the ALM values (vertical axis) at every iteration step (horizontal axis). The dashed lines indicate that the ALM is eventually constant. We observe that the trends displayed by each graph are consistent: the parameter $\lambda^{(k)}$ returned at the first iterations increases, then it descreases until it  settles down, with ALMA always stopping in a finite-time. This demonstrates its stability and effectiveness in achieving accurate image reconstructions across noise levels and undersampling rates. Subfigure B shows histograms of the number of iterations needed for ALMA's convergence across undersampling rates while subfigure C shows the same for different noise levels (see the legends). The number of bins for each histogram is the ceiling of the square root of the number of points, i.e., $\lceil \sqrt{150}\rceil=13$. We observe that the number of iterations for ALMA to stop is mostly concentrated in the interval $[0,10]$ independently of the noise level and the undersampling rate, while the standard deviations increase with the undersampling rate and the noise percentage. This demonstrates that the number of iterations required for ALMA to converge is relatively low even when processing larger amounts of information ($20\%$ sampling) and higher noise percentages ($7\%$ noise).}
	\label{fig:convergence}
    \end{center}
\end{figure*}
	
	For fixed noise level and undersampling rate, ALMA performs a reconstruction and updates the ALM approximation at each iteration. We have selected two stopping criteria:
	\begin{itemize}
		\item The number of iterations $k$ reaches a predefined maximum (set to $k_{max}=100$).
		\item At iteration $k_0+1$, $\lambda^{(k_0+1)}=\lambda^{(k_0)}$, in which case $\lambda^{(k)}=\lambda^{(k_0)}$ for every $k\geq k_0$.
	\end{itemize}
	Note that the second criterion serves as a convergence criterion for ALMA. This means that if it is satisfied, ALMA not only stops but actually \textit{converges} (i.e., the sequence $(\lambda^{(k)})_k$ of ALMs at step $k$ converges). We observe that in all of our experiments, the sequence of the ALMs is eventually constant. Therefore, in what follows, the term \textit{convergence} will refer to ALMA stopping due to the fulfillment of the second stopping criterion.
	
	Through empirical analysis, we investigated the convergence of ALMA across various noise levels and undersampling rates. As aforementioned,  ALMA converges in a finite time for every undersampling rate and noise level, fig. \ref{fig:convergence} (A). 

The histograms in fig. \ref{fig:convergence} illustrates the number of iterations required for the convergence of ALMA across different undersampling rates (B) and noise levels (C). Both histograms follow Gaussian distributions that become less concentrated around 0 as the undersampling rate increases (fig. \ref{fig:convergence} (B)) or as the noise level rises (fig. \ref{fig:convergence} (C)). In both scenarios, the increase in the number of iterations can be attributed to the amount of information processed by ALMA, which grows with the number of sampling points and the noise level. Let us delve deeper into these cases. On average, ALMA stops after $7.2089\pm2.9773$ iterations (see fig. \ref{fig:table1}). For the $10\%$ sampling rate, both the average number of iterations and the standard deviation increase with the noise level. A similar trend is observed for the $15\%$ undersampling rate, as detailed in Table \ref{fig:table1}. However, the $20\%$ sampling rate exhibits a distinct pattern: the average number of iterations increases from $3\%$ to $5\%$ noise levels and reaches its minimum at a noise level of $7\%$.

	\subsection{Performance of ALMA with respect to mSSIM}
    
\begin{figure*}
    \begin{center}
	\includegraphics[width=0.9\textwidth]{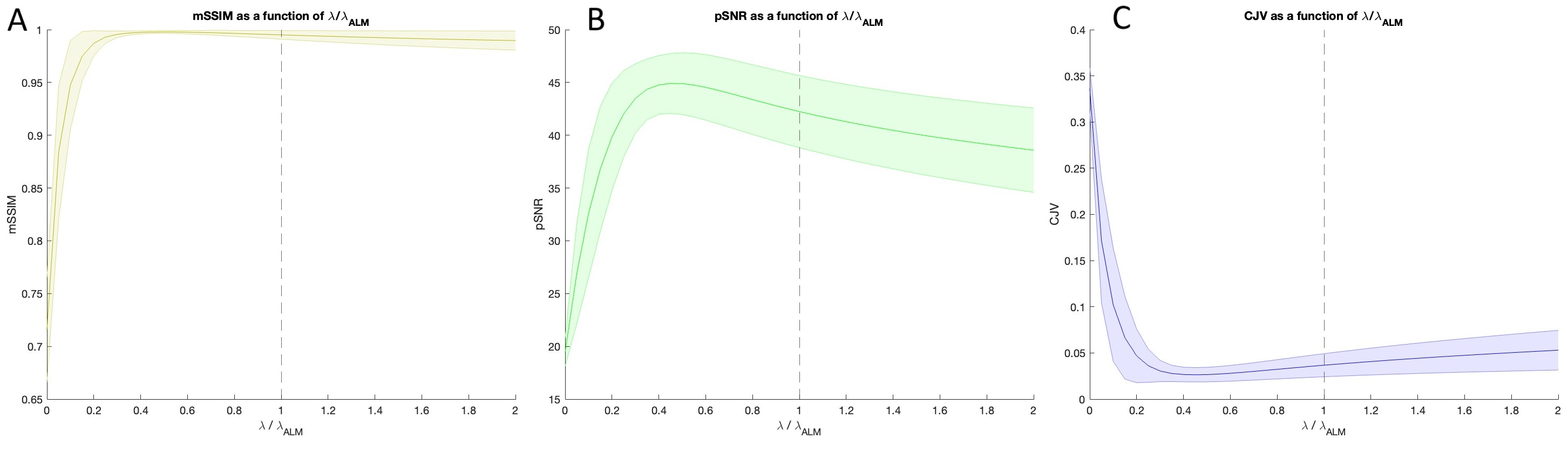}
	\caption{A: Shaded error bar of mSSIM as a function of $\lambda/\lambda_{ALM}$, the shade representing the corresponding standard deviations. The value 1 on the horizontal axis corresponds to $\lambda = \lambda_{\text{ALM}}$.We observe that the corresponding mSSIM values lie on the plateau of the mSSIM graph, just to the right of the maximum. This indicates that the ALM performs nearly optimally with respect to the mSSIM metric.
    B: Shaded error bar of pSNR as a function of $\lambda/\lambda_{ALM}$. The value 1 on the horizontal axis corresponds to $\lambda = \lambda_{\text{ALM}}$ and the shade represents the corresponding standard deviations. The corresponding points on the graphs are located to the right of the maxima, and this is more pronounced than in the mSSIM case. However, the pSNR of the reconstructions obtained using the ALMs remains high, indicating that the ALMs still produce high-quality reconstructions despite being slightly offset from the optimal point.
    C: Shaded error bar of CJV as a function of $\lambda/\lambda_{ALM}$. The value 1 on the horizontal axis corresponds to $\lambda = \lambda_{\text{ALM}}$, the shade representing the corresponding standard deviations. The corresponding points on the graphs are located to the right of the maxima, similar to the pSNR case. Nonetheless, the corresponding CJV values are low, indicating that the ALM also performs well with respect to the CJV metric.
    }
	\label{fig:SEB-mSSIM-pSNR-CJV}
    \end{center}
\end{figure*}

	Images reconstructed utilizing the ALMs computed by ALMA consistently exhibit an average $mSSIM \geq 0.99$, irrespective of the noise level or the undersampling rate (see figure \ref{fig:table1}). The violin plots for the mSSIM are displayed in figure \ref{fig:Tabellona} (color yellow).
 
Across noise levels and undersampling rates, the tuning parameter that optimizes the mSSIM ($\lambda_{mSSIM}$) is around half of the ALM ($\lambda_{ALM}$), i.e. $\lambda_{mSSIM}\approx 0.52\cdot \lambda_{ALM}$ (see figure \ref{fig:table1}). However, even if $\lambda_{ALM}$ does not maximize the mSSIM, the points $(1,mSSIM(1))$, which correspond to the mSSIM of the reconstructions obtained with ALMs, always lie on the plateau of the graph of the function $mSSIM(\lambda/\lambda_{ALM})$, indicating that $\lambda_{ALM}$ is in the range of tuning parameters corresponding to almost-optimal mSSIM values (see fig. \ref{fig:SEB-mSSIM-pSNR-CJV}). 

\subsection{Performance of ALMA with respect to pSNR}

	 Reconstructions obtained using ALMA exhibit high pSNR for every noise level and undersampling rate. As expected the pSNR decreases across noise levels. On average, the pSNR is $\geq 40$dB for reconstructions of images corresponding to $15\%$ and $20\%$ undersampling rates, whereas the average pSNR of reconstructions with $10\%$ undersampling rate is at least 35dB (see figure \ref{fig:table1}). Other than demonstrating the quality of reconstructions obtained via ALMA in terms of pSNR, this reflects the fact that solving TV-LASSO with ALMs as tuning parameters tend to produce highly regularized images. 
Again, across noise levels and undersampling rates, the tuning parameter that optimizes the pSNR ($\lambda_{pSNR}$) tends to be approximately 0.47 times the corresponding ALM, i.e. $\lambda_{pSNR}\approx 0.47\cdot\lambda_{ALM}$ (see figure \ref{fig:table1}). Compared to the graphs $mSSIM(\lambda/\lambda_{ALM})$, the plateau of the functions $pSNR(\lambda/\lambda_{pSNR})$ near their maxima is less pronounced, and the points $(1,pSNR(1))$, corresponding to the pSNR of reconstructions obtained with ALMs, are shifted to the right of their peaks (see fig.\ref{fig:SEB-mSSIM-pSNR-CJV}B). In conclusion, ALMs perform almost-optimally with respect to pSNR.

\subsection{Performance of ALMA with respect to CJV}

Except for the worst-case scenario $UR_\%=10\%$, $NL_\%=7\%$, the reconstructions obtained by ALMA display CJV values no larger than 0.05, showing that ALMA performs well with respect to CJV as well (see figure \ref{fig:table1}). 
The tuning parameter that minimizes the CJV ($\lambda_{CJV}$) is on average 0.45 times the corresponding ALM, i.e. $\lambda_{CJV}\approx0.45\cdot \lambda_{ALM}$ (see figure \ref{fig:table1}). However, ALMs still perform almost optimally with respect to CJV, and the points $(1,CJV(1))$, corresponding to the CJV of reconstructions obtained with ALMs, are shifted to the right of their minima, in accordance with the behavior of the graphs $pSNR(\lambda/\lambda_{CJV} )$ (see fig.\ref{fig:SEB-mSSIM-pSNR-CJV}C). 

\subsection{Comparison with the L-curve parameter}

For a comparison, we report on the quantitative measurement regarding the reconstructions obtained utilizing the parameter $\lambda_{L}$, computed by $L-$curve. These raconstructions exhibit an average $mSSIM \geq 0.99$, across noise levels and undersampling rates (see figure \ref{fig:table2}), they also exhibit an average pSNR of $42.6351$ and an average CJV of $0.035$. Therefore, they perform well regarding the metrics criterion. Moreover, the tuning parameter that minimizes the mSSIM ($\lambda_{mSSIM}$) is on average $0.6$ times the corresponding $\lambda_{L}$, i.e. $\lambda_{mSSIM}\approx0.6\cdot \lambda_{L}$, while the tuning parameters optimizing the pSNR and the CJV ($\lambda_{pSNR}$ and $\lambda_{CJV}$, resp.) are on average $0.5$ times $\lambda_L$ (see figure \ref{fig:table2}). All these data together show that the performance of the L-curve parameter is slightly better compared to the performance of the ALM. However, the difference is marginal and amounts to only a matter of decimals. Moreover, ALMA requires approximatively $7 \pm 3$ reconstructions in average (since a reconstruction occurs at each iteration) and in any case less that 30 reconstructions in our experiment, while the L-curve method required 41 reconstruction to reach the result exhibited here for comparison. Finally, the L-curve method is equivalent to an exhaustive searche and fails in principle if the lagrange multiplier to approximate is out of the bound of the exaustive searche. In contrast, ALMA do not require any apriori knowledge about the lagrange multiplier to estimate in order to succeed. All these facts taken together can be interpreted as a superiority of ALMA.

\begin{figure*}
\center
\includegraphics[width=0.98\textwidth]{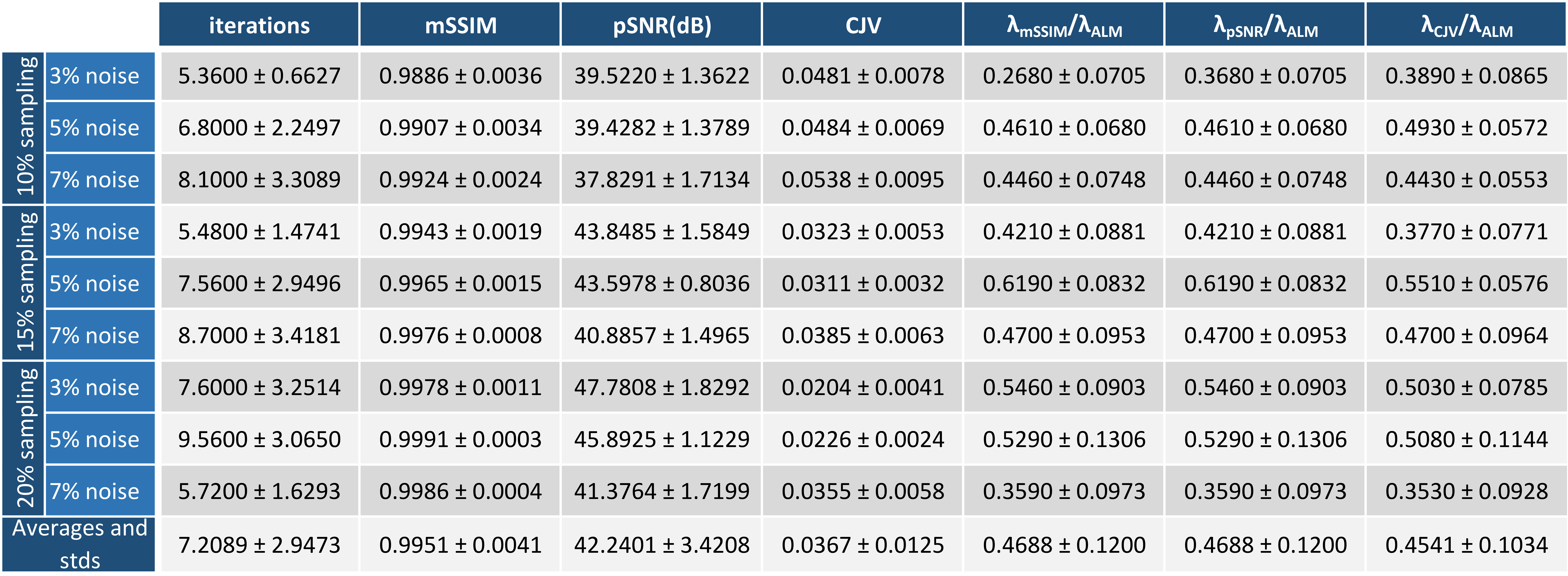}\\
\caption{Means and standard deviations of the three metrics, and the magnitude ratios across undersampling rate and noise level, for the reconstructions obtained using the ALM. Observe that the average number of iterations is 7, and the three metrics exhibit average values of approximately 0.9951, 42.2401 dB, and 0.0367, respectively, indicating optimal performance of ALMA. }
\label{fig:table1}
\end{figure*}

\begin{figure*}
\center
\includegraphics[width=0.98\textwidth]{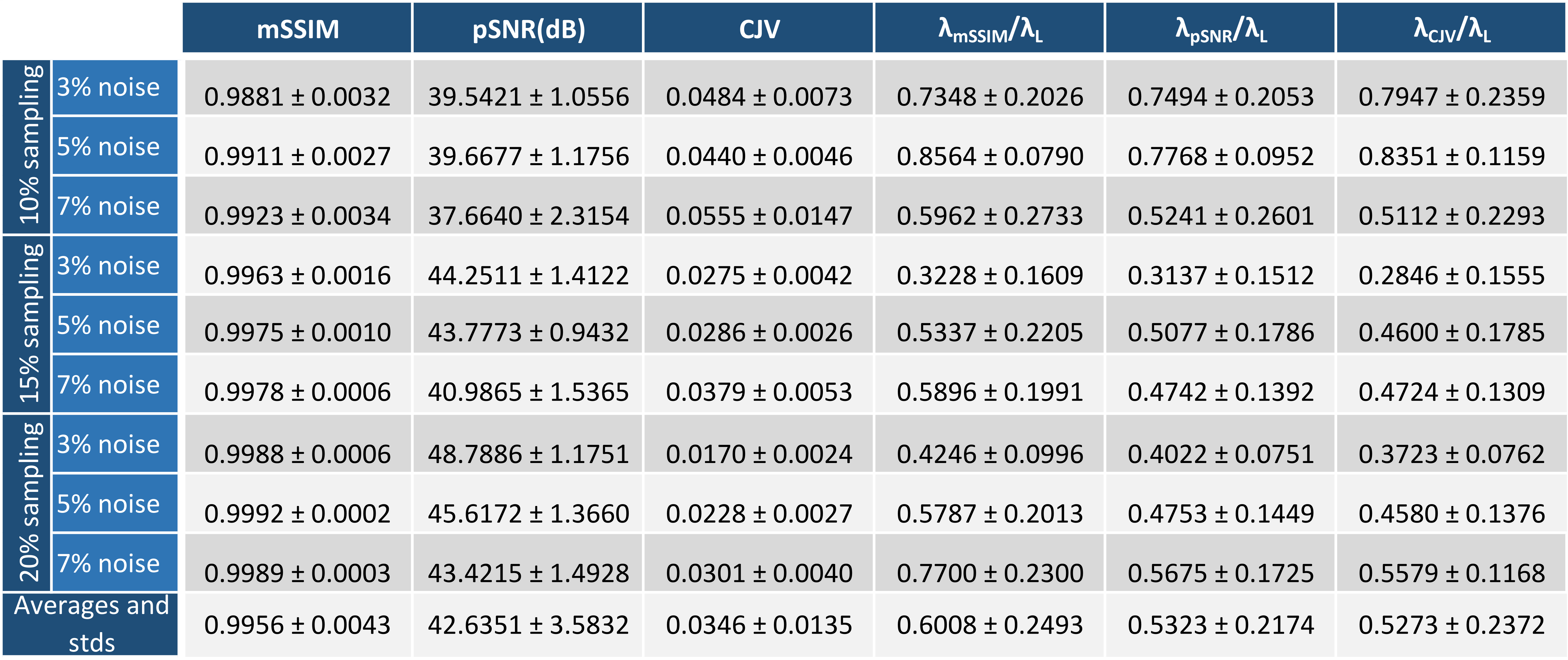}\\
\caption{Means and standard deviations of the number of iterations, the three metrics, and the magnitude ratios across undersampling rate and noise level, for the reconstructions obtained using $\lambda_L$, the tuning parameter of the L-curve.}
\label{fig:table2}
\end{figure*}

\section{Discussion}
We introduced ALMA for TV-LASSO, which demonstrated promising results in reconstructing undersampled and noisy MRI phantom data, achieving high-quality reconstructions without extensive manual tuning. We assessed the quality of the reconstructions quantitatively by means of mSSIM, pSNR and CJV. 

\begin{figure*}
\center
\includegraphics[width=0.5\textwidth]{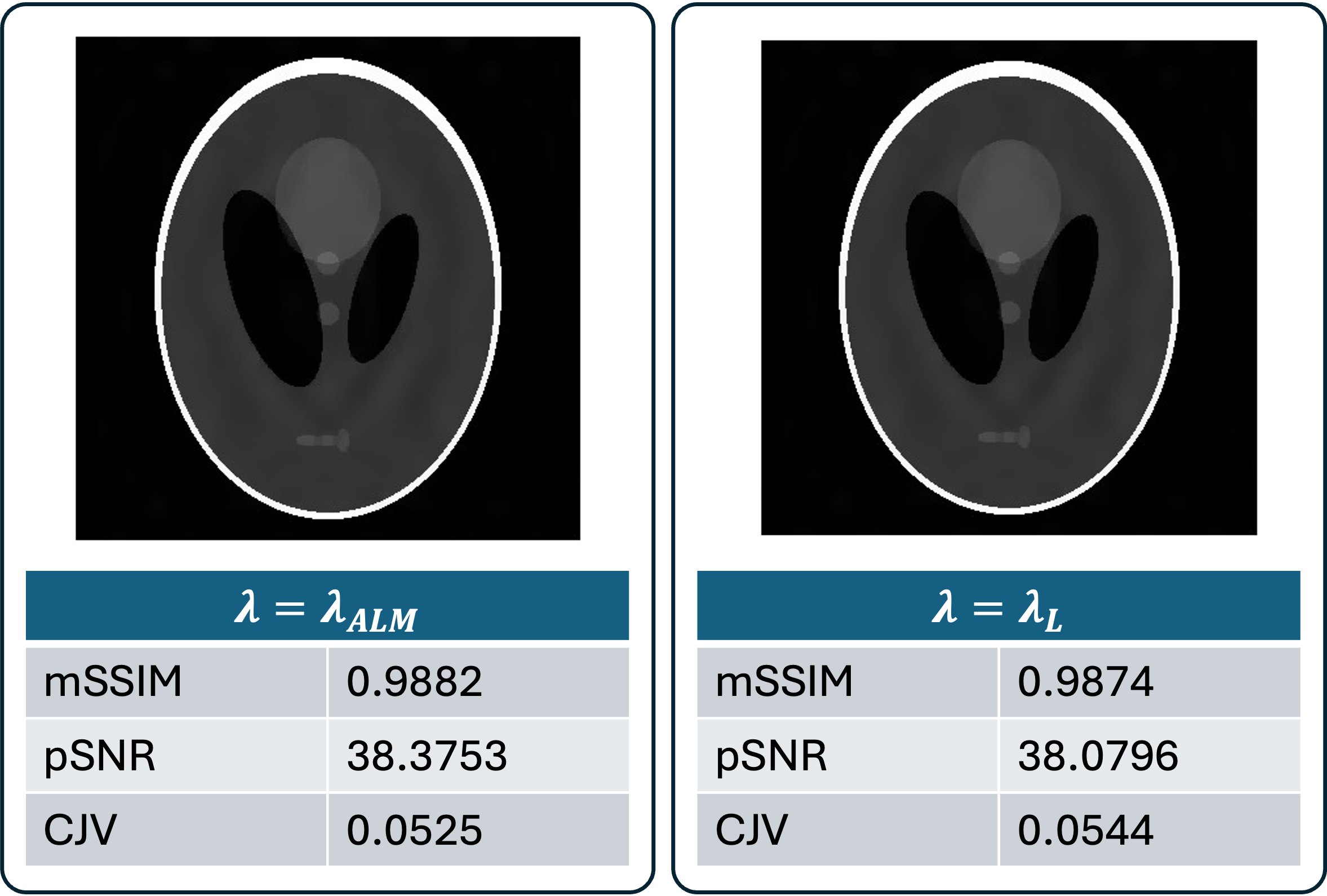}\includegraphics[width=0.5\textwidth]{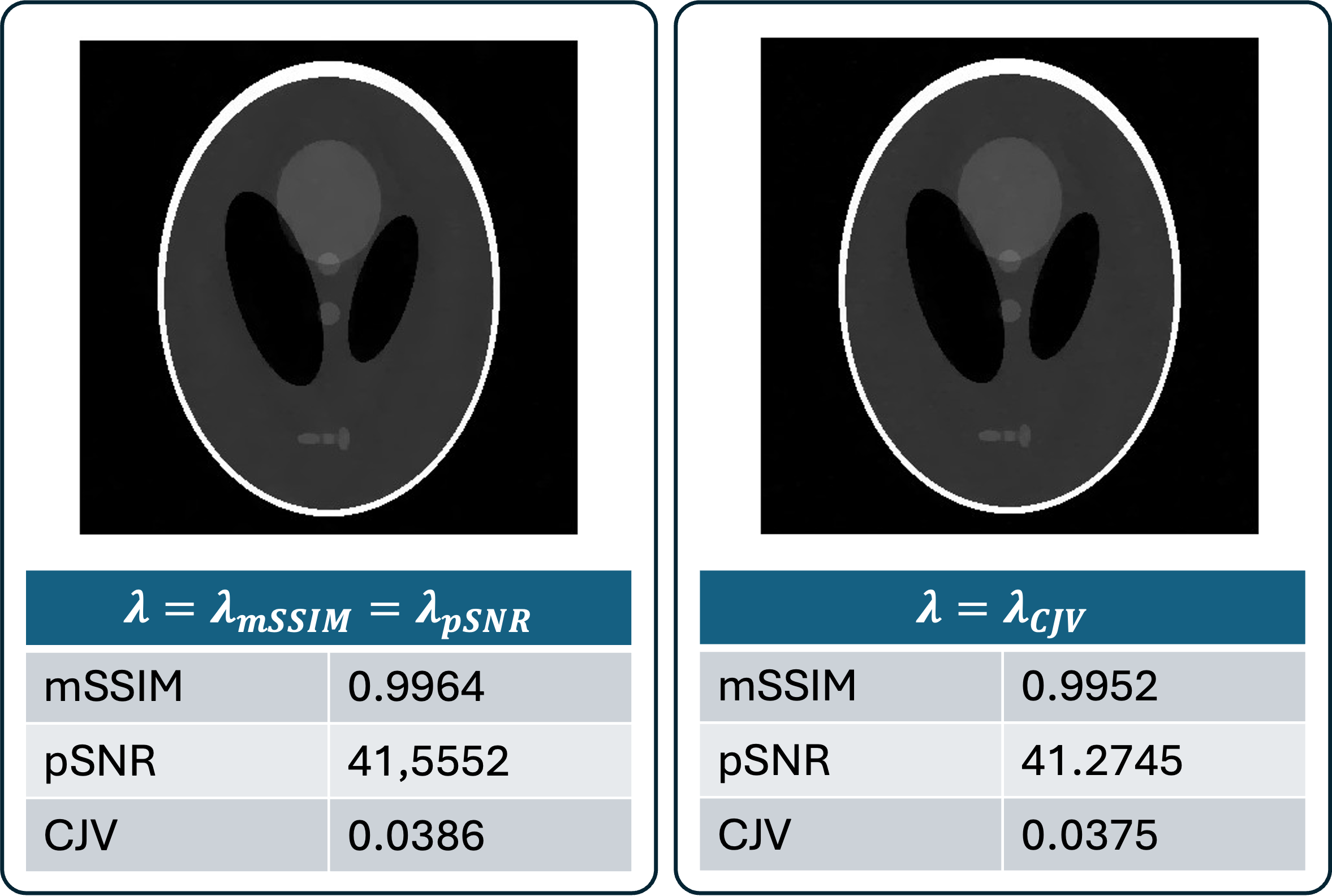}\\
\includegraphics[width=\textwidth]{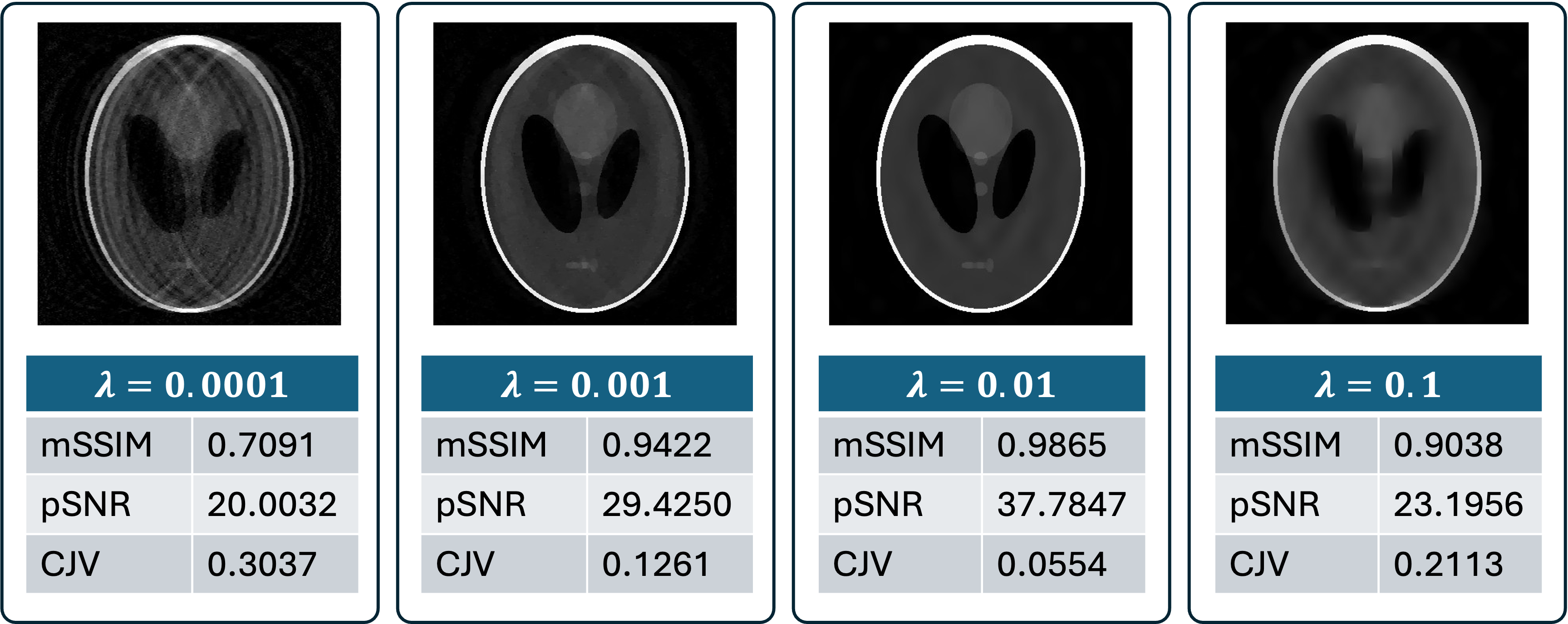}
\caption{Example of reconstruction of simulated MRI data ($NL_\%=7/100$, $UR_\%=10/100$) with different tuning parameters: the ALM and the tuning parameter computed with the L-curve method, the tuning parameters that optimize each metric and tuning parameters that are powers of 10. The corresponding quantitative quality assessments are reported below each reconstruction. We observe that the quality of the reconstructions obtained using the ALM and the L-curve parameter are nearly indistinguishable, and close to the quality of the images reconstructed with the tuning parameters that optimize the metrics.}
	\label{fig:ALMvsL}
\end{figure*}
		
As illustrated in fig. \ref{fig:ALMvsL}, the reconstruction quality using ALMA, even in the worst case scenario ($UR_\%=10\%$ and $NL_\%=7\%$), is notably superior when compared to parameter choices as powers of ten (see fig. \ref{fig:ALMvsL}). {The reconstructed image via ALMA exhibits finer structural details and improved contrast, which are crucial for accurate diagnosis and analysis in clinical radiology.}

To further validate ALMA's robustness, we compared its reconstructions to those obtained using parameters optimized for the three metrics (mSSIM, pSNR, and CJV). Figure \ref{fig:table1} encapsulates this comparison, while fig. \ref{fig:ALMvsL} provides a visual example of the reconstructions. The results indicate that ALMA's performance is nearly optimal, matching closely with the parameters that optimize each metric. 

We also compared the performance of ALMA against the well-established L-curve method. As presented in figures \ref{fig:table1} and  \ref{fig:table2}, and exemplified in fig. \ref{fig:ALMvsL}, both methods show comparable performance across the metrics. However, ALMA stands out as the first non-trained and self-calibrated algorithm that does not need any apriori knowledge on the lagrange multiplier to estimate and is also requires much fewer reconstructions in order to succeed, as compared to the L-curve method.  

Qualitatively, ALMA-reconstructed images maintain the anatomical integrity and structural fidelity necessary for clinical interpretation. This is evident in fig. \ref{fig:ALMvsL}, where even under severe noise and undersampling, the critical features are preserved. Quantitatively, ALMA achieves high mSSIM, pSNR, and low CJV values, closely approximating the optimal values for these metrics obtained through extensive parameter optimization.

The strength of ALMA, however, does not limit to TV based LASSO denoising, or to Cartesian undersampling. For example, we repeated our experimental framework in extreme conditions ($NL_\%=15/100$ and $UR_\%=15\%$), but with radial sampling. The quality of the reconstruction is undoubtedly comparable to the best reconstruction in terms of mSSIM, as displayed in fig. \ref{fig:radial}.

\begin{figure*}
\center	
\includegraphics[width=0.75\textwidth]{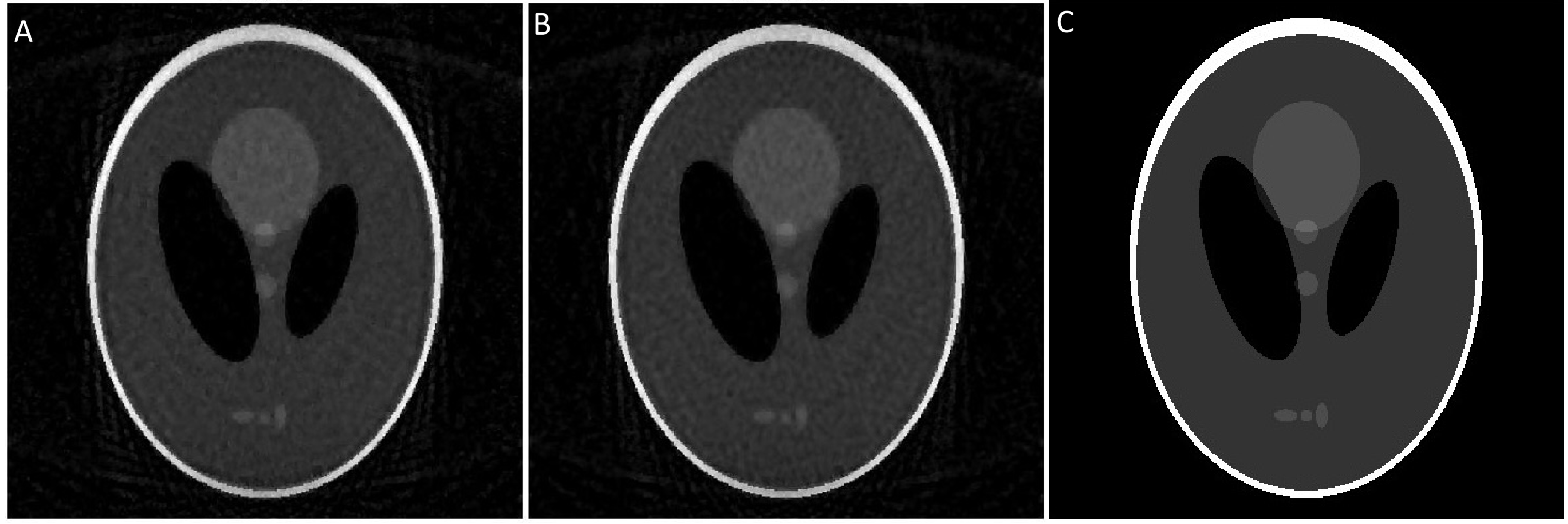}
\caption{Reconstructions of the Shepp-Logan brain phantom under radial sampling, with $NL_\%=15/100$ and $UR_\%=15\%$. The reconstruction obtained with the ALM is displayed in sub-figure A. The reconstruction obtained with the parameter that maximizes the mSSIM is displayed in sub-figure B. We observe that the two images are almost indistinguishable. Sub-figure C displays the ground-truth.}
	\label{fig:radial}
\end{figure*}

\section{Conclusions}\label{sec5}

We achieved our three main objectives:
\begin{itemize}
	\item We defined an iterative procedure to approximate Lagrange multipliers. 
	\item We demonstrated the efficiency of ALMs as tuning parameters for TV-weighted g-LASSO in the context of MRI.
	\item We assessed the quality of the reconstructions using image quality metrics, including mSSIM, pSNR, and CJV.
\end{itemize}
Our results show that ALMA performs almost optimally across varying levels of noise and undersampling, consistently yielding high-quality reconstructions in terms of image quality metrics. This iterative algorithm offers significant advantages by actively computing the tuning parameter during reconstruction, which enhances computational efficiency and ease of implementation. 

While our focus was on TV-LASSO, the principles underlying ALMA can be adapted to other models, paving the way for broader applications. For practical MRI applications, accurately estimating the norm of the noise remains a fundamental challenge for the implementation of ALMA. Such notwithdstanding, our work serves as a basis, providing a solid foundation for future research aimed at improving parameter estimation and enhancing MRI reconstruction techniques. 

\vspace{20pt}
\noindent
\large\textbf{ORCID} 

Gianluca Giacchi            : 0000-0002-6809-1311 

Isidoros Iakovidis          : 0009-0005-2594-0028 

Micah M. Murray             : 0000-0002-7821-117X 

Bastien Milani              : 0000-0002-2410-2744 

Benedetta Franceschiello    : 0000-0003-0754-5081

\vspace{20pt}
\noindent
\large\textbf{Funding Information} 

This work was partially supported by Swiss National Science Foundation; grant 220433 to BF and the PON scholarship PONDOT1303154-3

\vspace{20pt}
\noindent
\large\textbf{Author contributions} 

G.G. conducted the analysis, developed the mathematical framework, implemented ALMA and provided a first draft of the article. I.I. developed the comparison with the L-curve. B.M. provided the code implementation and contributed to the algorithm design. B.F. conceived the idea of the work, and together with B.M. supervised its development. All authors contributed to the final version of the paper. 

\vspace{20pt}
\noindent
\large\textbf{Financial disclosure} 

None reported.

\vspace{20pt}
\noindent
\large\textbf{Conflict of interest} 

The authors declare no potential conflict of interests.

\nocite{*}
\bibliographystyle{IEEEtran}

%\bibliography{MRM-AMA}

\end{document}